# Observation of Dirac hierarchy in three-dimensional acoustic topological insulators

Linyun Yang[1,#], Yin Wang[2,#], Yan Meng[1], Zhenxiao Zhu[1], Xiang Xi[1], Bei Yan[1], Shuxin Lin[1], Jingming Chen[1], Bin-jie Shi[2], Yong Ge[2], Shou-qi Yuan[2], Hong-xiang Sun[2,3*], Gui-Geng Liu[4,†], Yihao Yang[5,6,‡], Zhen Gao[1,§]

[1]Department of Electrical and Electronic Engineering, Southern University of Science and Technology, Shenzhen 518055, China.
[2]Research Center of Fluid Machinery Engineering and Technology, Faculty of Science, Jiangsu University, Zhenjiang 212013, China.
[3]State Key Laboratory of Acoustics, Institute of Acoustics, Chinese Academy of Sciences, Beijing 100190, China
[4]Division of Physics and Applied Physics, School of Physical and Mathematical Sciences, Nanyang Technological University, 21 Nanyang Link, Singapore 637371, Singapore
[5]Interdisciplinary Center for Quantum Information, State Key Lab. of Modern Optical Instrumentation, College of Information Science and Electronic Engineering, Zhejiang University, Hangzhou 310027, China
[6]ZJU-Hangzhou Global Science and Technology Innovation Center, Key Lab. of Advanced Micro/Nano Electronic Devices & Smart Systems of Zhejiang, ZJU-UIUC Institute, Zhejiang University, Hangzhou 310027, China
[#]These authors contributed equally: Linyun Yang and Yin Wang.

Dirac cones (DCs) play a pivotal role in various unique phenomena ranging from massless electrons in graphene to robust surface states in topological insulators (TIs). Recent studies have theoretically revealed a full Dirac hierarchy comprising an eightfold bulk DC, a fourfold surface DC, and a twofold hinge DC, associated with a hierarchy of topological phases including first-order to third-order three-dimensional (3D) topological insulators, using the same 3D base lattice. Here, we report the first experimental observation of the Dirac hierarchy in 3D acoustic TIs. Using acoustic measurements, we unambiguously reveal that lifting of multifold DCs in each hierarchy can induce two-dimensional (2D) topological surface states with a fourfold DC in a first-order 3D TI, one-dimensional (1D) topological hinge states with a twofold DC in a second-order 3D TI, and zero-dimensional (0D) topological corner states in a third-order 3D TI. Our work not only expands the fundamental research scope of Dirac physics, but also opens up a new route for multidimensional robust wave manipulation.

The groundbreaking discovery of Dirac cones (DCs) in graphene [1,2] and topological insulators (TIs) [3,4] has inspired extensive research toward Dirac physics to emulate hypothetical massless Dirac particles in distinct artificial systems, ranging from ultracold atoms [5,6] to photonic crystals [7–14] and phononic crystals [15,16]. The DCs in quantum or classical systems give rise to various intriguing phenomena, such as Klein tunneling [17,18], vanishing refractive index [19–22], and topologically robust transport [23–37]. Besides, via the introduction of mass terms to DCs, various topological effects such as the quantum Hall effect [38] and quantum valley/spin Hall effect [25–37] possibly arise.

In another context, three-dimensional (3D) TIs are a class of topological matters featuring 3D insulating bulk but conducting boundaries. According to the principle of bulk-boundary correspondence that was recently generalized by the newly-discovered high-order topology, an $n$th-order 3D TI has topologically robust states along (3-$n$) dimensional boundaries. For example, a first-order 3D TI has topological surface states with two-dimensional (2D) DCs in 3D bulk bandgaps; a second-order 3D TI has topological hinge states with one-dimensional (1D) DCs in 2D surface bandgaps; a third-order 3D TI has topological corner states in 1D hinge bandgaps. The 3D TIs thus provide a fertile field for studying Dirac physics. Although the 3D TI phases were initially proposed in condensed-matter systems, their realization in photonics and acoustics is currently under intense investigation, owning to their ability to wave trap and manipulation in a topologically robust manner across multiple dimensions, and their great potential in next-generation integrated photonics/acoustics. An important task in this field is to realize a full hierarchy of topological phases, i.e., first-order to third-order 3D TIs, using the same base configuration; in this way, robust 2D interfaces with surface DCs, 1D hinge waveguides with line DCs, and zero-dimensional (0D) cavities can exist in a single 3D integrated platform. However, all existing experimental works on 3D TIs are limited to a single or two topological phases [12,39–43]; a full topological phase hierarchy that consists of first-order, second-order and third-order 3D TIs based on the same configuration has remained elusive so far.

Recent theoretical advances have suggested a novel route to achieve the hierarchy of band topology of 3D TIs, which is intriguingly associated with a hierarchy of DCs [44] that corresponds to various topological phase transition points across multiple dimensions. 3D honeycomb lattices with the same base configuration can exhibit 3D bulk states with an eightfold DC, 2D surface states with a fourfold DC in a first-order 3D TI, 1D hinge states with twofold DC in a second-order 3D TI, and 0D corner states in a third-order 3D TI, which, however, have so far evaded experimental observation.

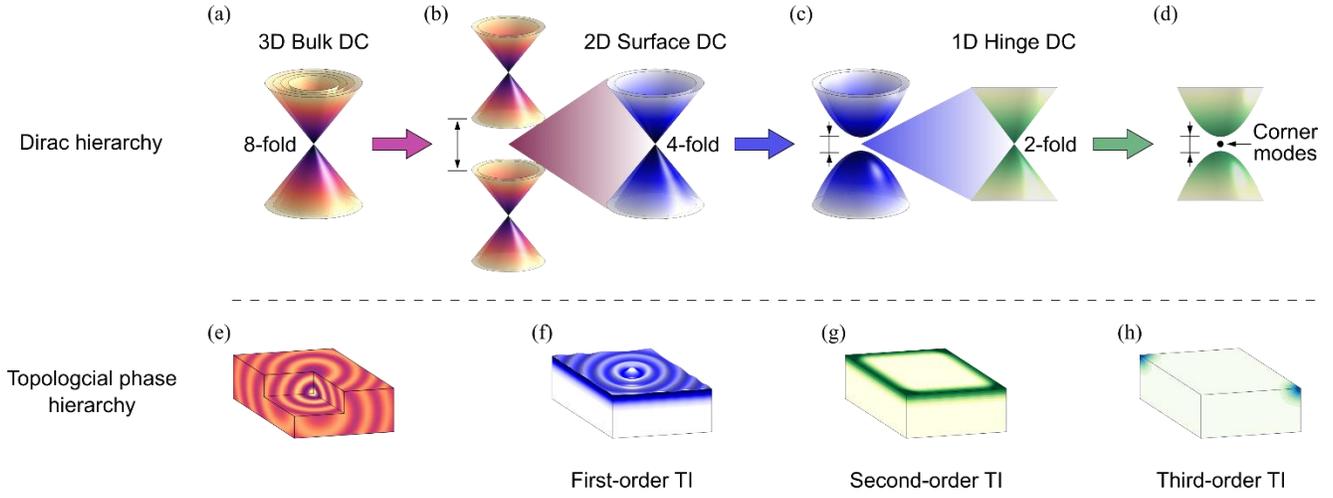

FIG. 1. Dirac hierarchy and topological phase hierarchy. The Dirac hierarchy comprises (a) an eightfold-degenerate bulk DC to (b) a fourfold-degenerate 2D surface DC in a first-order 3D TI, followed by (c) a twofold-degenerate 1D hinge DC in a second-order 3D TI and finally (d) a pair of 0D corner states in a third-order 3D TI. At each hierarchy, the degenerate DC in the higher dimension is broken by appropriate symmetry breaking, resulting in a complete bandgap that hosts a DC dispersion in a lower dimension. The topological phase hierarchy comprises of (e) a gapless phase, (f) a first-order TI, (g) a second-order TI, and (h) a third-order TI.

Here, we report unambiguous experimental evidence of the full hierarchy of DCs and the corresponding hierarchy of band topology in 3D acoustic crystals composed of stacked 2D honeycomb lattices of interconnected acoustic resonators. We start from a 3D acoustic crystal with an extended honeycomb cell containing multiple primitive unit cells, which features a 3D eightfold-degenerate bulk DC [see Fig. 1(a)] with 3D bulk states [see Fig. 1(e)]. By inducing alternating dimerization along the perpendicular direction, we can lift the eightfold bulk DC to create a first-order topological bandgap where the 2D topological surface states [see Fig. 1(f)] with a fourfold DC appear [see Fig. 1(b)]. By further inducing a Kekulé distortion in the horizontal plane, we can further lift the fourfold surface DC to create a second-order topological bandgap wherein the 1D hinge states [see Fig. 1(g)] with twofold DC exist [see Fig. 1(c)]. Finally, by breaking the in-plane mirror symmetry, we can lift the twofold hinge DC to create a third-order topological bandgap wherein the 0D corner states emerge [see Fig. 1(d) and (h)]. All these 3D TIs are characterized by in-plane and out-of-plane winding numbers. The full hierarchy of DCs and the topological phases has been directly visualized via the acoustic field imaging measurements and spatial Fourier transformation.

To experimentally demonstrate the Dirac and topological phases hierarchy, we fabricate a series of 3D acoustic crystals using a standard 3D printing technique. As shown in Fig. 2(a), the first experimental sample exhibits the first-order topological insulating phase, supporting 2D topological surface states with a fourfold DC. The details of the unit cell are shown in Fig. 2(b). It consists of two layers of honeycomb lattices in the horizontal plane ($xy$ plane) interconnected by vertical alternating air tubes with widths $r_s$ and $r_t$, respectively, forming a Su-Schrieffer-Heeger (SSH) chain. Within each layer, the Kekulé textured honeycomb lattice contains two types of air tubes with widths $r_w$ and $r_v$, respectively. See Supplementary Material for detailed parameters. Such an acoustic crystal can be mapped to a tight-binding model with a Hamiltonian

$$\mathcal{H} = I_2 \otimes \mathcal{H}_{xy} + \mathcal{H}_z \otimes I_6, \quad (1)$$

in which $I_N$ is a $N$-by-$N$ identity matrix ($N = 2, 6$), $\mathcal{H}_{xy}$ and $\mathcal{H}_z$ are the Hamiltonians for the monolayer Kekulé lattice and the out-of-plane SSH chain, respectively. Both of them feature block-off-diagonal forms that $\mathcal{H}_{xy} = [0, Q; Q^\dagger, 0]$ and $\mathcal{H}_z = [0, \rho; \rho^*, 0]$, in which $Q$ and $\rho$ are the couplings between different sublattices for the Kekulé lattice and SSH chain, respectively. For the explicit form of $Q$, $\rho$ and the Hamiltonian (1) under a specific basis, please refer to the Supplementary Material. After some algebra operation, one can find that the eigenvalue of $\mathcal{H}$ is $E = E_z + E_{xy}$, with the corresponding eigenstate $\psi = \psi_z \otimes \psi_{xy}$, in which $E_{xy(z)}$ and $\psi_{xy(z)}$ are the eigenvalue and eigenstate of $\mathcal{H}_{xy}$ ($\mathcal{H}_z$), respectively. This fact implies that: if $\psi_{xy}$ and are both bulk states, then $\psi$ is also a bulk state, while if $\psi_z$ is an edge state and $\psi_{xy}$ is a bulk (edge or 2D corner) state, then $\psi$ is a surface (hinge or 3D corner) state [44].

According to the zone-folding mechanism, when both interlayer couplings equal to each other ($r_w = r_v = 3.2$ mm) and the intercell coupling is equal to the intracell coupling ($r_s = r_t = 4.0$ mm), an eightfold bulk DC (red sphere) emerges at Z ($k_z = \pi$) in the band structure with the frequency around 4.6 kHz for the lowest twelves bands, as shown in Fig. 2(c).

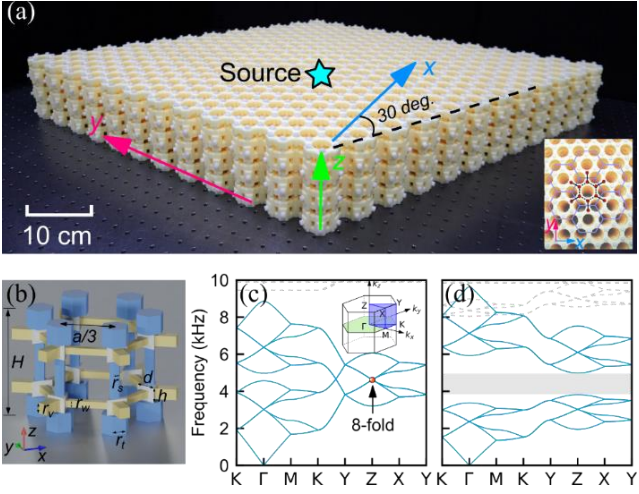

FIG. 2. Design and band diagram of the first-order 3D acoustic TI. (a) Photograph of the sample of the first-order 3D acoustic TI. The inset shows the top view of the sample. (b) Design of the unit cell of the acoustic crystal. The detailed geometrical parameters are provided in the supplementary materials. (c) Bulk band structure corresponding to equal out-of-plane and in-plane coupling strengths ($r_s = r_t, r_w = r_v$), hosting an eightfold-degenerate bulk DC at high symmetry point Z. (d) Bulk band structure for unequal out-of-plane coupling strengths and equal in-plane coupling strengths ($r_s < r_t, r_w = r_v$), in which case the eightfold-degenerate bulk DC is split into two fourfold-degenerate surface DCs with a complete 3D bandgap indicated by grey region.

By inducing an inequality in the interlayer couplings ($r_s =$ 2.0 mm, $r_t =$ 5.0 mm), the dimerization along $z$ direction lifts the eightfold DC degeneracy, leading to an extremely wide 3D topological bandgap from 3.8 kHz to 5.0 kHz (grey region), as shown in Fig. 2(d). Both band structures in Figs. 2(c-d) exhibit an approximately symmetric behavior about the frequency around 4.6 kHz and 4.4 kHz, respectively, which can be ascribed to the chiral symmetry of the SSH chain.

To demonstrate the first-order topological insulating phase and the 2D fourfold surface DC in the first acoustic crystal, we carry out a series of measurements. In the experiments, an acoustic point source [marked by a cyan star in Fig. 2(a)] is placed at the top center of the acoustic sample, which can generate a broadband signal from 2 kHz to 7 kHz to excite the surface and bulk modes simultaneously. A probe (microphone) is inserted into surface (bulk) resonators to measure the surface (bulk) transmission. The measured transmission along the surface (blue region) and in bulk (red region) are shown in Fig. 3(a). We observe a wide transmission dip from approximately 3.8 kHz to 5.0 kHz in the transmission of the bulk states, corresponding to an extremely wide bulk bandgap with more than 27 % relative bandwidth. On the surface, however, we observe the high transmission spanning the whole bandgap, indicating the existence of gapless topological surface states.

To map out the dispersion band structure of the topological surface states, we perform pump-probe surface measurements at the top $xy$ surface of the sample. The measured and simulated field distributions on the sample's external surface at 4.2 kHz are shown in Fig. 3(b), exhibiting good confinement of waves on the top surface. We then conduct spatial Fourier transform to the measured acoustic field distributions to obtain the surface dispersion. The measured surface dispersion along the high-symmetry line $\overline{K}$ - $\overline{\Gamma}$ - $\overline{M}$ - $\overline{K}$ (the projections of the high-symmetry points on the $xy$ plane) is displayed in Fig. 3(c), revealing a family of topological surface states spanning the frequency range of the bulk bandgap with a gapless fourfold DC around 4.3 kHz. The simulated surface dispersion (cyan hollow circles highlight the surface DC and transparent grey dots represent the bulk state) is also shown for comparison, matching well with the measured counterparts. We also plot the measured isofrequency contours in 2D reciprocal space from 3.9 kHz to 4.7 kHz, as shown in Fig. 3(d), confirming the conical nature of the surface dispersion. Moreover, we present in Fig. 3(d) the calculated isofrequency contours for comparison. The white dashed lines indicate the conical shape, and the double white solid circles in each isofrequency contour indicate the fourfold degeneracy of the surface DC that intersects at the surface Dirac point.

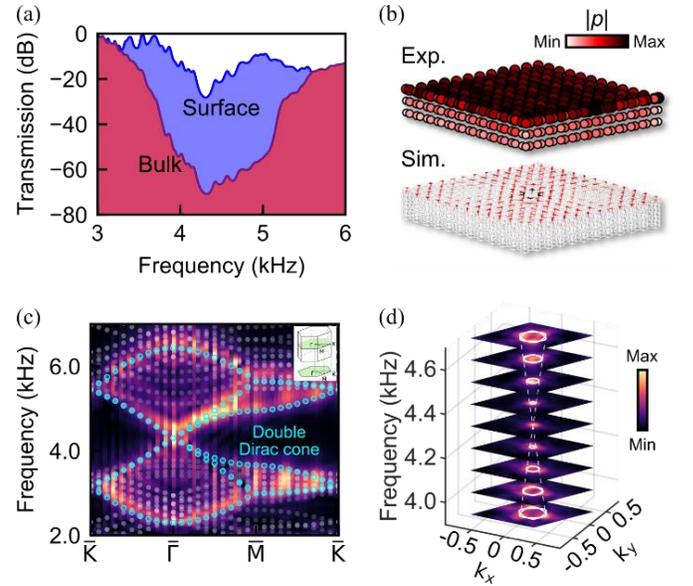

FIG. 3. Experimental observation of the fourfold surface DC in the first-order 3D acoustic TI. (a) Measured transmission spectra for the bulk (red region) and surface (blue region) states, respectively. (b) Measured and simulated acoustic field distributions of the topological surface states excited by a point source placed at the center of the top surface. (c) Measured (background colour) and simulated (cyan hollow circles and transparent grey dots) band diagrams of the topological surface states. The inset shows the projective Brillouin zone on the top surface. (d) Measured (background colour) and simulated (white circles) isofrequency contours of the topological surface states at different frequencies. White dashed lines are guides to the eyes that indicate the shape of the cones intersecting at the surface Dirac point, and the double white solid circles in each isofrequency contour indicate the fourfold-degeneracy of the surface DC. The colour scale measures the acoustic energy density.

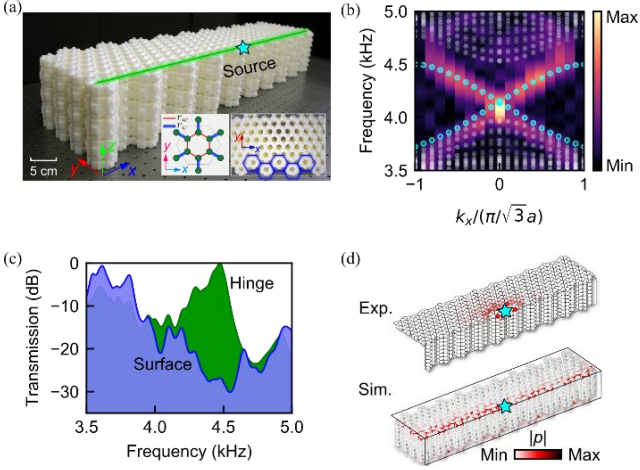

FIG. 4. Experimental observation of the twofold hinge DC in the second-order 3D acoustic TI. (a) Photograph of the sample of the second-order 3D acoustic TI. Left inset illustrates the in-plane Kekulé distortion with unequal intracell and intercell couplings. Right inset shows the molecule-zigzag hinge. (b) Measured (background colour) and simulated (cyan hollow circles) band diagram along the molecule-zigzag hinge direction ($x$ axis), which clearly exhibits a twofold-degenerate gapless hinge Dirac point. (c) Measured transmission spectra of the surface (blue region) and hinge (green region) states. (d) Measured and simulated acoustic pressure distribution of the hinge states excited by a point source (cyan star) placed at the middle of the hinge.

The topological nature of the first 3D acoustic crystal can be understood as follows. According to Eq. (1), the lattice configuration of the first sample indicates that $\mathcal{H}_{xy}$ has a gapless eigen-spectrum, and $\psi_{xy}$ is a bulk state, while $\mathcal{H}_z$ has a topological edge state $\psi_z$ because the winding number

$$w_z = -\frac{1}{2\pi}\int_0^{2\pi}\frac{d}{dk_z}\arg(\det\rho)\,dk_z \qquad (2)$$

is nonzero [44]. Consequently, $\psi = \psi_z\otimes\psi_{xy}$ appears to be 2D topological surface states.

Next, we experimentally demonstrate the second-order acoustic TI with twofold Dirac hinge states, by inducing in-plane Kekulé distortion to the first acoustic TIs to lift the fourfold surface DC. The second acoustic crystal sample with unequal intracell and intercell couplings ($r_w = 3.0$ mm, $r_v = 3.6$ mm, see the left inset of Fig. 4(a)) is shown in Fig. 4(a), which has a molecule-zigzag type hinge parallel to the $x$ axis [see the right inset of Fig. 4(a)]. Because of the in-plane Kekulé distortion, the fourfold degeneracy of the surface DC is lifted, and a complete second-order topological bandgap emerges approximately from 4.0 kHz to 4.6 kHz, wherein a pair of helical topological hinge states with twofold DC appear. As shown in Fig. 4(b), the measured (background colour) and simulated (cyan hollow circles) dispersion diagrams of the hinge state indeed reveal a gapless hinge Dirac point appears near the frequency of 4.2 kHz. We also measure the transmission spectra of the surface states (blue region) with a bandgap from 4.0 kHz to 4.6 kHz, but the transmission of the hinge state (green region) remains high with an enhancement of about 30 dB to that of the surface states, as shown in Fig. 4(c). To verify the hinge states are tightly confined and propagate on the hinge of the second-order acoustic TIs, we present in Fig. 4(d) the measured and simulated acoustic pressure distributions at 4.2 kHz. Both unambiguously show the sound is confined and guided along the hinge. Note that the discrepancy between the measured and simulated acoustic field distributions originates from the intrinsic absorption loss of the materials.

The emergence of these hinge states can be ascribed to a net-winding number when considering a specific lattice termination [44]. An alternative explanation can also be established by considering the mirror symmetry $M_y$ regarding the perpendicular direction of the molecule-zigzag hinge and calculating the mirror winding numbers. Since $M_y$ commutes with $\mathcal{H}_{xy}$ at Γ point, the off-diagonal block $Q$ can be block-diagonalized into two sectors $Q^\pm$, associated with the $+1$ and $-1$ eigenvalues of $M_y$ [45]. This further allows us to calculate the winding numbers in each sector separately as

$$w_{MZ}^\pm = -\frac{1}{2\pi}\int_0^{2\pi}\frac{d}{dk_\perp}\arg(\det Q^\pm)\,dk_\perp, \qquad (3)$$

in which the subscript denotes molecule zigzag edge, $k_\perp$ is the wave vector along the perpendicular direction of the edge. The calculated results show that $w_{MZ}^\pm = 0$ for $w > v$, while $w_{MZ}^\pm = \mp 1$ for $w < v$, which corresponds to the two helical hinge states in our case (see Supplementary Material). Moreover, the mirror symmetry $M_y$ necessitates these two helical hinge states degenerate at $k_x = 0$ [45], hence forming a 1D gapless hinge DC within the surface bandgap.

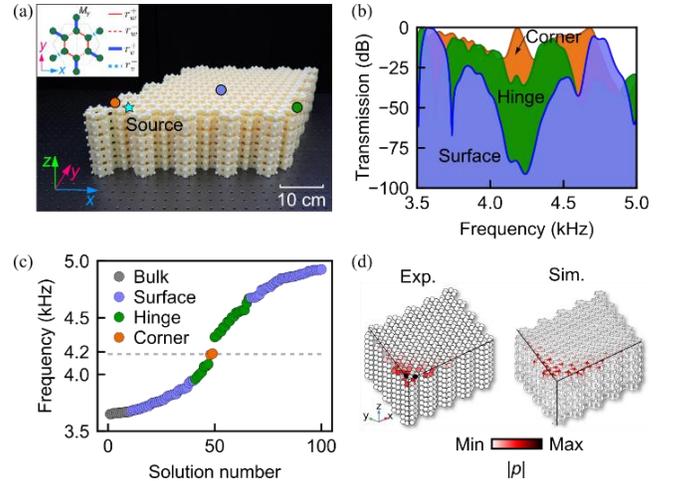

FIG. 5. Experimental observation of 0D corner states in the third-order 3D acoustic TI. (a) Sample of the third-order acoustic TI. Inset illustrates the $M_y$ mirror symmetry breaking. (b) Measured transmission spectra for the surface (blue region), hinge (green region) and corner (yellow region) states, respectively. (c) Calculated eigenfrequency spectra of the fabricated acoustic crystal. (d) Measured and simulated acoustic pressure distributions at the corner-

mode frequency (i.e., 4180 Hz) under a point source excitation placed near the sample corner.

Finally, we experimentally demonstrate the third-order 3D acoustic TI with 0D corner states. The third acoustic crystal sample has rhomb-shaped surface terminations with molecule-zigzag boundaries, as shown in Fig. 5 (a). The widths of the air tubes of the unit cell on the left side ($x < 0$) and on the right side ($x > 0$) are $r_w^- = 2.8$ mm, $r_v^- = 3.4$ mm and $r_w^+ = 3.1$ mm, $r_v^+ = 3.7$mm, respectively [see inset of Fig. 5(a)], which breaks the mirror symmetry $M_y$. We measure the transmission spectra of the surface (blue region), hinge (green region), and corner (yellow region) states under the excitation of a point source placed near a sample's corner. As shown in Fig. 5 (b), the corner measurement exhibits a transmission peak within the surface and hinge bandgap, which agrees well with the respective eigenfrequency ranges of the numerically calculated corner, hinge and surface eigenstates [Fig. 5(c)]. We also plot the measured and simulated acoustic pressure distributions at 4.18 kHz in Fig. 5(d). The resulting acoustic field distributions are indeed concentrated at the corner, revealing the localized characteristic of the topological corner state.

The appearance of the 0D topological corner states in the third acoustic crystal arises from the breaking of mirror symmetry $M_y$. This is because the twofold hinge DC in the second-order 3D acoustic TI is protected by the mirror symmetry $M_y$, breaking the mirror symmetry $M_y$ can lift the degeneracy of the hinge DC, resulting to the third-order topological phase.

We have thus experimentally demonstrated the full Dirac hierarchy across multiple dimensions with an eightfold-degenerate bulk DC, a fourfold-degenerate surface DC, and a twofold-degenerate hinge DC in 3D acoustic TIs using the same 3D base configuration. Moreover, a hierarchy of topological phases encompassing first-order TIs with 2D DC surface states, second-order TIs with 1D DC hinge states, and third-order TIs with 0D corner states are experimentally realized via step-by-step breaking different symmetries in a single acoustic 3D honeycomb lattice. Our work not only establishes a versatile platform for exploring exotic phenomena related to the hierarchy of Dirac physics and topological phases, but also opens up new avenues to achieve robust wave manipulation and trap devices across multiple dimensions, including 2D surfaces, 1D waveguides, and 0D cavities, in a single integrated photonic/acoustic circuit.

This work was sponsored by the National Natural Science Foundation of China under grant number 12104211, Young Thousand Talent Plan of China and the Southern University of Science and Technology for SUSTECH Start-up Grant (Y01236148, Y01236248). The work at Zhejiang University was sponsored by the National Natural Science Foundation of China under grant number 62175215. H.S. acknowledged the support of the National Key Research and Development Program of China (Grant No. 2020YFC1512403), and the National Natural Science Foundation of China (Grant No.12174159).

§gaoz@sustech.edu.cn,
‡yangyihao@zju.edu.cn
*jsdxshx@ujs.edu.cn
†guigeng001@e.ntu.edu.sg